\documentstyle[12pt,aasms4,rotate,psfig]{article}

\def\la{\mathrel{\mathpalette\fun <}}
\def\ga{\mathrel{\mathpalette\fun >}}
\def\fun#1#2{\lower3.6pt\vbox{\baselineskip0pt\lineskip.9pt
  \ialign{$\mathsurround=0pt#1\hfil##\hfil$\crcr#2\crcr\sim\crcr}}}
\lefthead{Ruderman et al.}
\righthead{Gamma-Ray Burst Sources}

\begin{document}

\title{A Central Engine for Cosmic Gamma-Ray Burst Sources}

\author{M.A. Ruderman\altaffilmark{1} and L. Tao} 
%W. Klu\'zniak\altaffilmark{1}}
\affil{Departments of Physics and Astronomy,\\
 Columbia University, New York, NY 10027}
% {\it email address:} mar@carmen.phys.columbia.edu}
\author{and\\ W. Klu\'zniak}
\affil{Copernicus Astronomical Center,\\
Warsaw, Poland}

\vskip 20pt

%\altaffiltext{1}{Copernicus Astronomical Center}
\altaffiltext{1}{Corresponding author. {\it E-mail address:} mar@carmen.phys.columbia.edu}

\begin{abstract}
One of a family previously proposed ``central engines'' 
for cosmic gamma-ray
burst sources (Klu\'zniak \& Ruderman 1998) 
is considered in some detail.  A steadily
accreting $10^6$ Gauss magnetic white dwarf should ultimately collapse
to a strongly differentially rotating, millisecond-rotation-period
neutron star for a wide range of steady accretion rates and initial
masses if the accreting white dwarf
 has an evolved O-Ne-Mg composition.  A similar neutron
star could also
result from an initial C-O white dwarf but only for more constrained
accretion rates.  Because the collapsing white dwarf begins as
a $\gamma=4/3$ polytrope, the final neutron star's spin-rate increases
strongly with cylindrical radius.  A stable wind-up of 
the neutron star's poloidal magnetic field
then produces buoyant magnetic toroids which grow, break loose, rise,
and partly penetrate the neutron star surface to form a transient,
$B\approx 10^{17}$
G millisecond spin-period pulsar with a powerful
pulsar wind (Usov 1992).  This
pulsar wind emission is then rapidly suppressed by the surface
shear motion from the strong stellar differential rotation.
This wind-up and transient pulsar formation
 can occur at other times on different cylinders and/or repeat
on the same one,
with (re-)wind up and surface penetration time scales hugely longer
than the neutron star's millisecond spin period.  
In this way, differential rotation both opens and closes the doors which
allow neutron star spin-energy to be emitted in powerful bursts
of pulsar wind.  Predictions of this model 
compare favorably to needed central engine properties 
of gamma-ray burst sources (total energy,
duration, sub-burst fluctuations and time scales, variability
among burst events, and baryon loading).
\end{abstract}


\section{Introduction}

Gamma-ray bursts (GRBs) are observed daily from sources at distances 
extending out to those of the oldest galaxies in our Universe.  To account 
for details of these bursts, ``central engines'' (CEs) of the GRB sources
should have the following properties (see Klu\'zniak \& Ruderman 1998,
hereafter KR,
for details and references).

(a) {\it Energy.}
Some CEs must store and release of order $10^{53}$ ergs (assuming
modest beaming of the energy outflow).

(b) {\it Fluctuations.}
There are often large temporal variations in the CE power output.  A CE
should be capable of attaining peak power within tens of milliseconds and
exhibiting large fluctuations thereafter.  The main power emission
is often in 
sub-bursts between which the CE is relatively dormant,
typically for about $10$ seconds, but sometimes for as long as several
$10^2$ seconds or as short as $10^{-1}$ seconds.

(c) CE {\it lifetimes}, typically seconds to tens of seconds, extend from 
less than a second to greater than $10^3$ seconds.  (There is also some
indication of an association of greater total energy release with longer
CE lifetimes.)

(d) {\it Baryon loading.}
The energy released from the CE of a GRB source carries with it at
most only a tiny baryon load of mass $\la 10^{-4} M_\odot$.

(e) The {\it birth rate} of GRB sources $\ga 10^{-6}$/galaxy/yr
(see, for example, B\"ottcher \& Dermer 2000).

(f) There is a very great variability among observed GRB events: durations,
time scales within a burst, and pulse shape structures, sub-burst numbers,
etc., vary so much that one cannot really specify a typical GRB.

The shortest time scales of (b) together with the total energy emission 
(a) suggest a CE formation involving stellar collapse to a neutron star
or to a black hole, or a very tight binary of such collapsed 
objects, or as part of some exotic supernova which would form such
objects.  However, the lifetimes (c), baryon loading (d), 
the commonly observed repeated widely separated  fluctuations
(b), and perhaps the birthrate (e) may raise special problems for such
CE models.  Particularly significant is why, if the CEs are collapsed
objects whose periods of rotation and vibration are expected to be 
milliseconds, energy emission from them so often involves several 
timescales which can be up
to $10^6$ times longer.

A promising way of constructing CE models based upon collapsed objects,
which incorporates this needed family of relatively long time scales,
begins by converting the most of the released collapse energy
into rotational energy of the collapsed objects.  The subsequent transfer
of that energy to emitted power in a form useful for ultimate 
$\gamma$-ray production may then be accomplished relatively slowly.
It is generally necessary to have CE magnetic fields 
$B\ga10^{15}$ G to extract the rotationally stored energy fast enough.
Such a CE model was long ago proposed by Usov (1992).
A millisecond spin-period
pulsar with a magnetic field $B \approx 10^{15}$ G was assumed to
be formed from an accretion induced collapse of a strongly magnetized
($B \approx 10^{9}$ G) white dwarf.  This simple CE model would be expected
to have the needed energy (a), lifetime (c) and baryon loading (d)
properties, but a sufficiently high birthrate (e) may be questionable
and the required fluctuation property (b) does not seem to be realized.

It has been proposed more recently that very large differential rotation
plays an essential role in CE models (KR).  One
such model has significant similarities to Usov's proposed 
millisecond-period
``magnetars'', but the initial white dwarf's pre-collapse history
and magnetic field strengths differ, and there are essential
differences in what happens within the neutron star and on
its surface. This strongly differentially
rotating CE would form and evolve in
the following, quite different, way.

1) A common ``garden-variety'' magnetic white dwarf ($B \approx 10^6$ G)
in a tight binary is spun up to its 
equilibrium spin-period ($P \approx 10^3$ s)
by an accretion disk fed by its companion.

2) The accreting white dwarf is either an evolved one (O-Ne-Mg), or a
canonical (C-O) dwarf, with accretion rates such that the accreting
white dwarf increases its mass, implodes before its growing stellar
mass reaches $1.4 M_\odot$, and collapses to a neutron star.

3) A neutron star is then formed with an initial spin-period $P \approx
10^{-3}$ seconds, a nearly canonical pulsar polar magnetic surface
dipole component
$B_p \approx 10^{12}$ G, and, most importantly, a spin-rate which 
increases very greatly with distances from the star's spin-axis.  It is
this crucial last feature which is the reason for choosing here to
discuss this particular CE model from among the previously suggested
possibilities for CEs with large initial differential rotation
(KR).

4) An interior toroidal field ($B_\phi$) is then stably wound up from 
the poloidal field ($B_p$) by this differential rotation until 
$B_\phi \approx 10^{17}$ G.  After that $B_\phi$ is achieved, the
wound-up (and probably slightly twisted) 
toroid's magnetic buoyancy for the
first time exceeds interior anti-buoyancy forces (from compositional
stratification).
 The buoyant toroid pushes up to the surface by moving
parallel to the spin-axis up to, and then partly penetrating the
stellar surface, within about $10^{-2}$ seconds after its initial
release.

5) For as long as some of this magnetic field sticks out of the rapidly
spinning neutron star's surface, this will be an extreme 
realization of
an Usov pulsar, a hyper-magnetar powered by the star's spin energy.
It is, however, extremely transient because of surface movements.

6) This surface dipole field (and higher multipoles) can survive
for only a very brief time ($\approx 10^{-2}$ seconds): it is continually
smeared out around the spin-axis,
and thus diminished by the strong 
on-going differential rotation
shearing the surface below any protruding field.  (There may also be
considerable surface field reconnection after this.)

7) After the first break-out of wound-up toroidal field, 
surface penetration by some of it, and
the resulting transient Usov pulsar, a similar
wind-up of the $B_p \approx 10^{12}$
G may begin again as in Step 4, around the same cylinder or a
somewhat slower wind-up may exist on some other cylinder).
In tiehr case, a new toroid grows until its $B_\phi$ reaches 
$B_\phi(max) \approx 10^{17}$ G when another sub-burst occurs as in
Steps 5-6.  The characteristic interval between the first 
and second sub-burst would be
\begin{equation}
\tau_{sb} \approx \frac{2\pi B_\phi(max)}{(\Delta\Omega) B_p} \approx
10 \, \, {\rm seconds},
\label{eq:subbursttime}
\end{equation}
where $\Delta\Omega$ is the spin-frequency difference between the
inner and outer parts of the differentially rotating neutron star.

8) The GRB source's CE finally turns off completely when either of two
stages is reached by the engine:

a) the differential rotation ($\Delta\Omega$)
which drives the wind-up of $B_\phi$ becomes
so diminished by the conversion of the differential rotation energy into
toroidal field energy that it can no longer cause build-up to the critical
$B_\phi\approx 10^{17}$ G needed for a pulsar wind sub-burst, or

b) the stellar spin ($\Omega$ of the outer region) becomes so reduced in the
transient pulsar phases sustained by it that pulsar wind
emission is almost extinguished even if a huge protruding field were 
to survive.

In this present note, we consider the above GRB source CE proposal in
more detail and discuss why and how it should have all of the 
desired properties.

\section{Accretion Induced Collapse of Magnetic White Dwarfs to Neutron
Stars}

Some white dwarfs (WDs) 
in tight binaries can accrete enough mass from their companions
to initiate implosions (because of electron capture by nuclei)
as they approach (but just before they reach)
their Chandrasekhar limits.
After such an implosion begins, there is a competition between energy
release from nuclear fusion reactions which act to explode the star and
a growing rate of electron capture---which removes pressure support and
accelerates collapse.  The winner in this competition, which depends
upon these relative rates, determines whether such WDs end
as Type Ia supernovae (no remnant star) or as neutron stars.  Figure 1
shows how the ultimate fate of such accreting WDs is determined
by the mass of the WD when accretion begins ($M$) and the
steady accretion rate ($\dot M$) which brings it to the initial implosion
instability (typically when the accreting stellar mass is about
 $1.35 M_\odot$). 
There are three possibilities. 1) The accreting WD has $M$ and
$\dot M$ in the cross-hatched region. Then nova explosions continually
eject at least as much mass as is accreted between these nova explosions
and the implosion mass is not reached.  
2) The WD's $M$ and $\dot M$ begin in the unmarked
region.  It then ends its life by an accretion induced collapse (AIC) to a
neutron star.  3) When $M$ and $\dot M$ are in the dotted region, the 
accreting WD ends in an explosion with no remnant---a Type Ia
supernova (SN)---if its composition is initially C+O.  If it approaches
implosion with a more evolved O+Ne+Mg composition, sustained $\dot M$
ultimately causes it to collapse to a neutron star (Nomoto \& Kondo
1991; see also Bailyn \& Grindlay 1990).

(Below we shall consider WDs with magnetic fields $B \approx 10^6$ G
and, mainly because of that magnetic field, 
spinning with periods $P\approx 10^3$ seconds.  Neither are of
much consequence in the early stages of collapse of these WDs
when the ultimate fate of the WD is determined.  The magnetic field
energy density is approximately $10^{-9}$ that of the WD's rotational
kinetic energy, which, in turn, is 
approximately $10^{-3}$ that of its gravitational
binding energy.  Therefore Figure 1 should not be sensitive to a
WD's possible $10^6$ G field or $10^3$ second spin period.)

A WD with a $B \approx 10^6$ G has a relatively modest field among
the ``magnetic white dwarfs'' in the local WD population.
Based upon those, they might be expected to number several percent
of the WD population.  Because of this field, an accretion
disk around such a WD, fed by mass pulled from its companion, should
spin up the accreting WD to a steady state angular spin rate
\begin{equation}
\Omega \approx \frac{\dot M^{3/7} (G M)^{5/7}}{(BR^3)^{6/7}}
\approx {\dot M_{18}}^{3/7} 10^{-2}\, \, {\rm seconds}^{-1},
\label{spinrate}
\end{equation}
with $R$ the WD radius and $\dot M_{18}$ the WD accretion rate in
$10^{18}$grams/second ($\dot M_{18}=1$ when $\dot M = 2\times
10^{-8} M_\odot$ per year).
For equation (\ref{spinrate}) to hold, it is assumed that $\dot M$
is small enough to keep the inner edge of the accretion disk above
the stellar surface, i.e., 
\begin{equation}
\dot M < \left(\frac{R^5 B^2}{G M}\right)^{1/2} \approx
10^{20}  {\rm g/s} = 2 \times 10^{-6} M_\odot \, \, {\rm year}^{-1}
\end{equation}
The total mass which would have to be accreted to reach the $\Omega$
of equation (\ref{spinrate}) is about $10^{-2} M_\odot$.  Thus before
magnetic WDs with a dipole field
$B\approx 10^6$ G accrete enough to collapse,
it is reasonable to expect a good fraction of them, probably most, to have
been spun-up to a period $P_{WD}\approx 10^3$ seconds.  After they
have collapsed to neutron stars with $R\approx 10^6$ cm, those neutron
stars would then have 
\begin{equation}
P_{NS} \approx 10^{-3}\, \, {\rm seconds.}
\label{PNS}
\end{equation}

If magnetic flux is conserved during the collapse, a very plausible
approximation because of the short time for collapse 
(on the order of seconds),
these millisecond period neutron stars are formed with (poloidal)
fluxes
\begin{equation}
B_{NS} \approx 10^{12}\, \, {\rm G} 
\label{BNS}
\end{equation}
If such neutron stars are to be candidates for GRB source CE's, their
formation rate must be $\ga 10^{-6}$yr$^{-1}$-galaxy$^{-1}$.  
Type Ia supernovae are observed to occur at a rate 
$2\times10^{-3}$yr$^{-1}$-galaxy$^{-1}$.  A plausible guess for the
fraction of $B\approx10^6$ G exploding WDs among them $\approx 2\times
10^{-2}$, if this fraction is about the same as that for such fields
to be found in the local WD population.  The fraction of such moderately
magnetized WDs in cataclysmic variables 
(accreting WDs in tight binaries) is
very much greater than this.  (However, their number
statistics are subject to significant but still unquantified selection
effects.)  A fraction of $2\times10^{-2}$ assumed above, for
WDs which become Type Ia SNe, seems
rather conservative based on present knowledge about these WDs. 
Then if more than only $3\times10^{-2}$ of the
WDs which accrete enough to explode as Type Ia SNe had a 
composition, or a combination of initial $M$ and $\dot M$, to implode
to neutron stars, the formation rate for neutron stars satisfying
equation (\ref{PNS}) and equation (\ref{BNS}) would be enough
for them to be a candidate population for CEs if the other required
properties are met.

The simplest of these and most necessary to satisfy is the
maximum energy requirement (a).
The spin-energy of a neutron star with an average rotation rate
($\bar \Omega \approx 10^4/$second)  about
$\approx 10^{53}$ ergs is difficult
to compare precisely with CE requirements because of the still
unknown GRB beaming and some uncertainties in the neutron
star equation of state and moment of inertia.  We turn next to
other special properties of these particular AIC formed neutron
stars which determine CE fluctuation timescales (b), lifetimes (c) and
baryon loading (d) and variability among the family
of these engines (e).

\section{Initial Differential Rotation of the Neutron Star}

The pressure support in a WD whose mass approaches $1.4 M_\odot$
is from extreme relativistic electrons; the star is 
a $\gamma=4/3$ polytrope.  Such a star has a central density ($\rho_c$)
very strongly peaked relative to its average density $\bar \rho$:
$\rho_c\approx 55 \bar \rho$ (Shapiro \& Teukolsky 1983). 
The difference in $\Omega$ between the inner
and outer parts of the newly formed neutron star will depend on the
initial composition of the imploding WD.  A C-O WD, when collapse
begins from $^{16}O + e^- \rightarrow ^{16}C + \nu$, has $\rho_c\approx
2\times 10^{10}$g/cm$^3$.  A O-Ne-Mg WD, whose collapse is
initiated by $^{24}Mg + e^-\rightarrow ^{24}Na + \nu$, has $\rho_c\approx
3\times 10^9$g/cm$^3$.  If during collapse to a 
neutron star the angular momentum
were to be conserved independently in each of the ``rings'' of matter
circulating around the spin-axis, then the
final spin-rates of rings which were originally in from the central
region of the collapsing $\gamma=4/3$ WD are much less than those
of rings collapsing in the outer regions.
Roughly, the average neutron star spin
\begin{equation}
\bar\Omega_{NS} \approx\Omega_{WD} \times
\left(\frac{\bar\rho_{NS}}{\bar\rho_{WD}}\right)^\frac{2}{3}
\approx 10^4/{\rm second},
\end{equation}
where $\bar\rho_{WD}$ is the initial average WD density and
$\bar\rho_{NS}$ that of the neutron star.
However, for the central region of the WD
\begin{equation}
\rho_c(WD) \approx 55 \bar\rho_{WD},
\end{equation}
compared to the very much more modest peaking for the central region of
a $1.3 M_\odot$ neutron star,
\begin{equation}
\rho_c(NS) \approx 5 \bar\rho_{NS}.
\end{equation}
Insofar as the pressure support of a 
somewhat cooled neutron star can be approximated as that of a
non-relativistic neutron kinetic energy ($\gamma=4/3$), a neutron
star's $\bar\rho/\rho_c\approx 6$ (Shapiro \& Teukolsky 1983).
Additional contributions to stiffening the neutron star's 
equation of state  (which must be present to increase its maximum
neutron star mass from the Oppenheimer-Volkoff $0.7 M_\odot$
to the observed range which is at least twice as large)
reduce this ratio further.
Therefore the central regions of this newly born neutron star
should initially be spinning much less rapidly than most of the 
matter in that star by a factor of about
\begin{equation}
\left[\frac{\bar\rho_{NS} \, \, \rho_c(WD)}
{\bar\rho_{WD}\, \, \rho_c(NS)}\right]^{-2/3} \approx
10^{-2/3} \approx 0.2\, \, ,
\end{equation}
but, other than the fact that this number is very considerably
less than $1$, its precise value will not be important in the
approximations considered below.

To the approximation that the pressure in the neustrop stellar matter 
depends only
on density, a dynamically stable steady state is finally
achieved  after fluid flow adjustments give
an $\Omega_{NS}$ which depends only
on the distance from the spin axis ($r_\perp$) (the
Taylor-Proudman theorem). In this idealization, the newly formed neutron
stars rotates on cylinders whose angular speed, $\Omega(r_\perp)$,
increases strongly with increasing $r_\perp$ because of
the differences in density distribution between
 the $\gamma=4/3$ polytrope
WD and the neutron star to which it implodes.
A crucial question is whether
this differential rotation might first have been
dissipated during collapse and, if it has survived, what becomes of
it in the next $10^3$ seconds or so.

During collapse, canonical viscous coupling between distant parts
of the star (e.g., by Ekman pumping) is far too weak to be important.
However, exchanges of angular momentum by transient energetic neutrino
transfer needs special consideration.  If this were important,
it would be expected to be most efficient 
for neutrinos whose mean-free path ($\lambda$) is
of order $R_{NS}\approx 10^6$ cm. These are emitted in
canonical SNEs from the rapidly
cooling neutron star remnants (e.g. the neutrinos detected from the SN 1987A
explosion) mainly over a 10
second interval.  In the implosion leading to the special model CE
neutron stars of interest here, most of the released energy should go
into stellar rotation rather than thermal heating.  Thermal
neutrino emission during and just after neutron
star formation should, therefore,
be much less.  If we approximate angular momentum transfers from 
neutrino transport
by assuming, say, a single absorption or scatter before escape (i.e.,
$\lambda \approx R_{NS}$), then the ratio of angular momentum transfer
by emitted neutrinos of total energy ${\cal E}_\nu$ to the total angular
momentum of the neutron star would be $\approx {\cal E}_\nu/Mc^2$.
Because the difference in angular momentum between the inner and
outer parts of the CE neutron star is comparable to the entire
stellar angular momentum, the fraction of that difference which
would be dissipated before neutrino cooling would also be of
order ${\cal E}_\nu/Mc^2$.  This ratio is certainly less than $10^{-1}$
and may be much less.

Another concern is the possibility of convective overturn 
(Ardeljan et al. 1996),
 which would mix (Taylor-Proudman) 
cylinders rotating with different angular
speeds.  (Fluid movements perpendicular to $r_\perp$ are not relevant.)
But on the short timescales of interest here, where viscosity and
thermal conduction are negligible, this should be strongly suppressed by
the great increase of angular
momentum per unit mass with increasing $r_\perp$:
$\frac{\partial}{\partial r_\perp}(\Omega^2\rho r_\perp)$ is much greater than any plausible 
convective force density when 
$\Delta\Omega\approx\Omega\approx 10^4$/second and 
$kT\la 10$ MeV.

During WD collapse there is also a small transfer of angular momentum
between different
collapsing regions by magnetic fields which couple them.
The WD's (polar) magnetic field which connects differently spinning
rings during the collapse would take a time 
\begin{equation}
\tau_A \approx
\frac{(4\pi)^{1/2} R_{WD}}{B_{WD}}\left(\frac{\bar\rho_{WD}}{\rho}\right)^{1/2}
\approx 10^6 \left(\frac{\bar\rho_{WD}}{\rho}\right)^{1/2} \, \, {\rm
seconds}\end{equation} 
to transfer angular momentum between them, where $R_{WD}$, $\bar\rho_{WD}$
and $B_{WD}$ are the WD radius, density and magnetic field
at the beginning of the
collapse, and $\rho$ is the (transient) 
density at any stage of the collapse.
This $\tau_A$ is far too long for the magnetic threading during
collapse to be a concern in modifying differential rotation.

This leaves the one mechanism for short time scale
dissipation of differential
rotation which is fundamental to our model for the CEs of GRB sources.
Because the differentially rotating cylinders of the newly formed neutron
star are coupled by the polar magnetic field ($B_p$) 
in the stellar interior,
that field will begin to wind up a toroidal one, $B_\phi$.
We turn next to
the stability, magnitude, and termination of that wind-up.

\section{Stability of Toroidal Field Wind-up}

The initial $\Omega(r_\perp)$ in neutron stars
formed in this particular AIC genesis
is one which grows strongly
with increasing $r_\perp$:
\begin{equation}
\frac{\partial}{\partial r_\perp} \Omega > 0 \, \, .
\label{diffrot}
\end{equation}
This rotating fluid is certainly very
linearly stable to axisymmetric hydrodynamic perturbations,
according to the Rayleigh criterion that the angular
momentum $r_\perp^2\Omega$ increases outwards.   Such a
neutron star is also not unstable (or, at least, no instabilities
have been discovered) to the 
so-called Tayler instabilities (Tayler 1973; see also Spruit 1999),
which do exist in the special case when 
$\partial \Omega/\partial r_\perp = 0$.
i.e., when the star is rotating rigidly. 
However, important recent work in MHD stability
theory has shown that powerful instabilities may exist 
in differentially rotating systems when they contain even
relatively weak magnetic fields (Velikhov 1959, Chandresekhar 1961, 
Balbus \& Hawley 1991).
Could the differential rotation of a neutron
star satisfying equation (\ref{diffrot}) be
unstable to any of these magneto-rotational instabilities?
Demonstrations of related MHD instabilities in differentially 
rotating objects have included non-axisymmetric perturbations,
compressibility, and toroidal
and poloidal fields
(see, for example, Balbus \& Hawley 1992 and
Ogilvie \& Pringle 1996).  Magneto-rotational
instabilities have been found
when angular velocity decreases with $r_\perp$ but no
instabilities have been exhibited for flows satisfying 
equation (\ref{diffrot}).  Indeed that inequality adds
to stabilizing forces in those cases for which the opposite
inequality causes instability.
Therefore it seems plausible that strong differential rotation
satisfying equation (\ref{diffrot}) will be stable 
even when the differentially spinning object is threaded by
a weak magnetic field: it will wind up an
initial poloidal field, which threads it, 
into a toroidal field until that field becomes
unstable because of buoyancy effects, or if magnetic 
forces grow to exceed gravitational ones.

For the magnetic toroid to become buoyantly unstable, the buoyant forces
must overcome whatever anti-buoyant stratification may exist in the
star.
(For a cold neutron star, 
the stratification
is the compositional one from varying proton/neutron
ratios, which can adjust via $n\leftrightarrow p+e$ with
neutrino emission being too slow to be efficient here.)  
The wound-up toroidal magnetic field
$B_\phi$ would be stable until the buoyancy force density
\begin{equation}F_b \approx \frac{B_\phi^2}{8\pi c_s^2}g\, \, ,\end{equation}
where $c_s$ the speed of sound of the embedding medium,
exceeds any anti-buoyancy force density.
Wind-up would ultimately increase $B_\phi$ until it reaches
a critical value $B_b$ at which the buoyancy force is enough
to balance the neutron star's interior anti-buoyancy.
This has been estimated to give $B_b$ of order $10^{17}$ G
(KR).  (Our estimates in this paper
do not depend upon knowing this $B_b$ accurately.  It is
certainly less than the equipartition value for a neutron
star whose gravitational binding energy $\approx 10^{-1} M_{NS} c^2$:
$B_\phi(equipartition)\approx 10^{18}$ G.  The KR
estimate of $B_b\approx 2\times 10^{17}$ G, based solely upon
anti-buoyancy forces from compositional stratification in a cooled
neutron star, varies only as the square root of that force and
is not changed qualitatively by inclusion of other, mainly
thermal, contributions).

Thereafter, a (probably
slightly twisted) toroid should
rapidly rise towards the stellar surface,
moving in the direction aligned with both the spin-axis $\vec \Omega$
and $\partial
{\vec\Omega}/\partial r_\perp$.
The wind-up of the non-axisymmetric
poloidal field into a strong toroidal component
may introduce some twist into the overwhelming toroidal field.
(For the released toroid to rise stably through the neutron
star, some twist may be what keeps it from fragmentation by
Kelvin-Helmholtz or Rayleigh-Taylor instabilities; see Tsinganos 1980).
It is appropriate to emphasize that we do not have a detailed
description of the wound-up toroidal bundle's dynamical evolution
after it is released by buoyancy.  The buoyancy force can push
up the fluid column above it (acting like a plug) to make a surface
bulge which can spread horizontally, and/or the approximate axial
symmetry of the wound-up toroid may be diminished by some
magnetized fluid movement perpendicular to $\vec r_\perp$.

\section{Surface Field Penetration: Transient Pulsar 
(Hyper-Magnetar) Formation}

As this toroid rises, the toroidal magnetic flux continues
to increase because it is still being wound up from the poloidal
component by the continuing differential rotation. 
As a function of time, $t$, measured from the moment 
when the field first reaches the critical strength $B_b$,
\begin{equation}
B_\phi = B_b + t B_p \Delta\Omega \, \, ,
\end{equation}
where $\Delta\Omega$ is again the characteristic difference in angular
velocity across the wind-up region.  (In our model the initial
 $\Delta\Omega\approx\Omega$.)
The buoyance force density, which depends on the square of
the toroidal component ($B_\phi$) minus that of the
anti-buoyancy force density which balances it when $B_\phi=B_b$, 
grows nearly linearly with $t$:
\begin{equation}
F_b \approx \frac{t B_b B_p \Delta\Omega}{4\pi c_s^2} g\, \, .
\end{equation}
Then the buoyancy timescale for rising up to and 
penetrating through the stellar
surface is
\begin{equation}
\tau_b \approx \left(\frac{24 \pi R c_s^2\rho}
{B_b B_p g \Delta\Omega}\right)^{1/3}.
\label{tau_b}
\end{equation}
For the special AIC formed neutron star,
the radius $R\approx10^6$ cm, the speed of
sound $c_s\approx 10^{10}$cm/sec, $B_b\approx 10^{17}$ G,
$B_p\approx 10^{12}$ G, $g\approx 10^{14}$cm/sec$^2$, and $\Delta\Omega
\approx 10^{4}$/second. Then the buoyancy, post break-free,
time scale of equation (\ref{tau_b})
is of the order $10^{-2}$ seconds.  At the beginning
of the wind up, the non-axisymmetric $B_p$ is not negligible relative
to $B_\phi$.  Again during the interval $0 < t < \tau_b\approx 10^{-2}$
seconds when the (positive) difference between the buoyancy and 
anti-buoyancy forces is small, non-axisymmetric forces from $B_p$ may
not be entirely negligible compared to other axisymmetric ones acting
on the toroids.  One effect of this would be a tilt to
the rising wound-up toroids accomplished by slightly different
forces and fluid movements in directions aligned with $\vec\Omega$
(and $\partial\vec\Omega/\partial r_\perp$).
Because the toroid will not be exactly axisymmetric, a part of it
will ultimately poke
through the surface of the star.
The field that penetrates the stellar surface will
 not escape because of the huge conducting mass
threaded by the rest of the toroid still below the surface
to which it is still strongly attached.
This strongly conducting mass 
remains gravitationally bound to the star.

Therefore $\tau_b$ of equation (\ref{tau_b}) 
is the turn-on time for the star becoming a Usov type (hyper-magnetar)
pulsar with a magnetic dipole field less than, but probably comparable
to $B_\phi \approx 10^{17}$ G.  (Because $\Omega R/c \approx 1/3$,
more than just the dipole component of this pulsar may be important
in analyzing its properties.)

%Therefore a toroid of $10^{17}$ G field rises to the surface of
%the star in about one hundredth of a second.  The field will then
%penetrate the surface and expand as we depict in Figure 3.  So
%long as this field sticks out of the surface of this rapidly
%spinning neutron star, the star will be a pulsar. 

\section{Transient Pulsar Termination: Extinction
of Surface Multipoles by the Surface Shear from Differential
Rotation}

Because of the continuing differential rotation of the star and
stellar surface, a surface dipole (and all other 
multipole) cannot survive long
beyond the characteristic differential
rotation periods involved.  If $\tau_{s}$ is
the time it take for the differential rotation to shear out the
surface dipole (multipole),
the characteristic value of the transient surface
field would be
\begin{eqnarray}
B_{surf} \approx B_b \times \frac{\tau_{s}}{\tau_b} &\, \, &{\rm if} \, \,
\tau_{s} < \tau_b,\cr
B_{surf} \approx B_b  &\, \, & {\rm if} \, \,
\tau_{s} > \tau_b.
\end{eqnarray}

%For a field such as that depicted in Figure 3, the shear destruction time of
%the protruding surface field is
%\begin{equation}
%t_{s} \approx \frac{2\pi r_1}{a_1} \frac{1}{\frac{\partial \Omega}
%{\partial r}{a_1}} \approx \frac{2\pi}{\Omega}\left(\frac{r_1}{a_1}\right)^2
%\approx 10^{-2} \, \, {\rm seconds.}
%\end{equation}

Figure 2 shows a simplified example of effects of 
surface shear motion suppression
of surface field moments beginning from a north-polar cap of
radius $a$ at a distance $d$ from the spin-axis $\vec\Omega$
together with a similar south-polar cap 
displaced by the same $d$.
Because of the different angular velocities of the
different parts in both polar caps (increasing with $r_\perp$),
the caps will be deformed into tighter and tighter
interwoven spirals extending from $r_\perp = d - a$ to $d+a$.
After many relative wind-ups between $r_\perp = d - a$ and
$r_\perp = d + a$, the surface field will be entirely smeared out
and cancelled.  If the cap radii or distances differed slightly,
very little of the surface field would survive as rings around
the spin-axis leading to hugely 
reduced power in the pulsar wind. [If $\epsilon$ is
a measure of the small difference in the two polar cap radii ($a$)
or their distances from the spin-axis ($d$),
then the asymptotic power
output after extended shearing
is reduced by a factor of order $\left(\frac{\epsilon}{a}\right)
\left(\frac{\epsilon\Omega}{c}\right)^2
\approx 10^{-2}$
relative to that from the initial configuration.]

The timescale for this surface dipole suppression to be essentially
completed is
\begin{equation}
\tau_{s} \approx {\rm several}\times\frac{2\pi}{\frac{\partial\Omega}
{\partial r_\perp}{a}} \approx {\rm several}\times
\frac{2\pi}{\Omega}\left(\frac{R}{a}\right)
\approx 10^{-2} \, \, {\rm seconds,}
\end{equation}
where we assume $\partial \Omega/\partial r_\perp\approx \Omega/R\approx
2\pi/10^{-3} R$ seconds, and several$\times R/a\approx 10$.
This estimated $\tau_s$ should be characteristic for the suppression
of all important surface multipoles of typical surface field geometries.

Therefore a $10^{17}$ G toroid rises to the surface of the star
in $\tau_b\approx 10^{-2}$ seconds, partly penetrates that
surface and expands outside the star.  It
survives for a time $\tau_{s}$ which is about
$10^{-2}$ seconds. % for typical CE parameters.
So long as this field
sticks out of the surface of this rapidly spinning neutron star, 
the star will
have the canonical spin-down power and wind emission of a pulsar
with dipole field $B_d\la 10^{17}$ G and $P\approx 10^{-3}$ seconds.

During this time, the field will be not only be smearing out into a ring
but also the north and south poles of the field will be brought into
closer and closer contact with each other.  This can result in some
reconnection.
However, reconnection is not the
dominant process of field
destruction and facilitates field destruction only
after the field is already smeared out.  

\section{Sub-burst structures: Energies, Intervals and CE durations}

During that brief interval while
the millisecond
pulsar's $B_d\approx 10^{17}$ G dipole field exists, 
its pulsar wind (consisting of
electromagnetic energy, $e^{\pm}$, and some baryons) carries away a
sub-burst energy
\begin{equation}
{\cal E}_{sb} \approx \frac{B_d^2 R^6 \Omega^4}{c^3}\times \tau_{s}
\approx 10^{52}\, \, {\rm ergs}
\label{e_sb}
\end{equation}
(where it is assumed that
the dipole $B_d\approx B_b$).  This is about that required for sub-bursts
in observed GRB events.  After the wound-up toroidal field breaks away
(and penetrates the surface), wind-up by still existing $B_p$
could continue to give yet another such wind-up of $B_\phi$
to $B_b$, and another transient pulsar, the rewind-up interval, i.e.,
the time between sub-bursts,
\begin{equation}
\tau_{sb} \approx \left(\frac{2\pi}{\Delta\Omega}\right)
\frac{B_b}{B_p}\approx \frac{10\, \, {\rm seconds}}{(B_p)_{12}},
\label{tau_sb}
\end{equation}
where $(B_p)_{12}$ is the poloidal field strength in units of $10^{12}$ Gauss.

In addition and perhaps even more important,
depending upon the details of $B_p$, there would often be significant
simultaneous winding up at other rates
in several different cylindrical regions
within the star.  Then sub-bursts from the transient pulsar formation 
which is the result of the wind-ups would
occur from them at different times and could vary enormously 
with a typical separation of about
the $\tau_{sb}$ of equation (\ref{tau_sb})
but often much longer or shorter than that.

These sequences of toroidal field wind-up, breaking free, surface 
penetration of magnetic field $B_d$, transient hyper-magnetar
pulsar wind, and
its suppression can continue only as long as $\Delta\Omega$
remains large enough to sustain yet another wind-up of $B_\phi$ to
$B_b$.  Typically the energy in the remaining differential rotation
within the neutron star would then have to be 
greater than about $(1/10) B_b^2 R^3
\approx 10^{51}$ ergs, typically $10^{-2}-10^{-1}$  times the equipartition
energy.

The ultimate energy source for the sub-bursts of pulsar wind emission
is the spin-energy of the entire star, the differential rotation serving
only as the key to open the pulsar wind
emission gate for brief intervals.  Because
this total ($E_{max}$) has a maximum of about  $I\bar\Omega^2/2
\approx 10^{53}$ ergs, it could sustain repeated strong emission
activity for a characteristic time (cf. equation (\ref{e_sb}))
\begin{equation}
\tau \approx \frac{E_{max}}{{\cal E}_{sb}} \times \tau_{sb}
\approx \frac{10^2\, \, {\rm seconds}}{(B_p)_{12}}.
\end{equation}

Some CE's may have a $\Delta\Omega$ which could give rise only to a
single sub-burst with a $\tau_b\approx 10^{-2}$ second rise time,
but the turn-off time ($\tau_{s}$) of such a CE should be stretched
considerably for small $\Delta\Omega$.  It is difficult to compare
sub-burst turn-off times directly
 with observations of the emission of $\gamma$-rays
powered by the pulsar wind sub-burst because these observed $\gamma$-rays
are created in relativistic expanding regions so very from from the CE.

\section{Baryon loading and beaming}

The transient emission (sub-bursts) from a CE is not what is directly
observed in GRBs.  The emission radius of observed $\gamma$-rays must
not be less than about $10^{15}$ cm if $\gamma+\gamma\rightarrow e^++e^-$
is not to absorb those $\gamma$-rays far above the
pair creation threshold.  At these large
radii, almost all 
the CE emitted burst energy has all been transferred 
by expansion into kinetic
energy of its co-moving baryon load.  To account for the short time scale
of many sub-bursts ($\approx$ a second), the observed emitting region must
be expanding relativistically with such a large Lorentz $\gamma$ that
the rest mass of baryons $\la 10^{-4} M_\odot$ for each $10^{53}$ ergs
in bursts.

If, as indicated in the previous section, the sub-burst emission is
powered by wind from a pulsar with $\Omega\approx 10^4$/second and
transient dipole $B\la 10^{17}$ G, the maximum possible baryon
outflow from the pulsar should be the maximum for a canonical
pulsar with these parameters.  The flow rate of nuclei with
charge $Ze$ out
from the stellar surface ($\dot N_Z$) 
should then not exceed the Goldreich-Julian
limit which would quench that outflow:
\begin{equation}\dot N_z \la \frac{\Omega^2 B R^3}{e c}
\approx 10^{-15} B_{17} M_\odot/{\rm second}.\end{equation}
This baryon load is negligible in itself and also compared to what the
pulsar wind, and especially the first sub-burst, would sweep up from
matter around the WD beyond the pulsar.

As noted in Section 6, 
there may also be small contributions to CE 
sub-bursts from some reconnection of magnetic loops
which extend up from the stellar surface.  While these loops are
essentially free of baryon loading 
above the surface (beyond the negligible Goldreich-Julian
one), how much they might pull out and up with them
during reconnection is far less clear.  Of course, each emission 
sub-burst from the pulsar need not itself be a source of power for
ultimate $\gamma$-ray emission. Some might become beam dumps
for slightly faster, later, much higher Lorentz $\gamma$
pulsar-wind sub-burst emissions with much less
baryon loading.

If most of the CE emission is in transient pulsar winds from a spinning
ephemeral surface dipole (or higher multipoles) as described above,
that dipole is mainly
orthogonal to the neutron star spin $\vec \Omega$.
In the simplest models for pulsar wind emission, 
with only electromagnetic power in the wind
from the star, the emission
 would be proportional to $\cos^2\theta$ with $\theta$
 the emission angle with respect to the spin axis.  Beaming in the
spin direction would then be a modest $3$ times the emission's
angular average.  Higher multipoles ($\Omega R/c \approx 1/3$) could
significantly increase this beaming.

\section{Variability among GRB events}

Details of emissions from these proposed GRB source CEs should be
sensitive to initial properties of the imploding ancestral WDs. There is an
important dependence on the ancestral WD's $\dot M$ and, 
especially, details of its magnetic field. 

(a) If the accreting WD's dipole component $B$ is much less than
$10^6$ G, its steady state spin period
from equation (\ref{PNS}) becomes so small
that it would not quickly and simply form a nuclear density
neutron star.  Centrifugal forces would stop much of that collapse
before it had evolved that far (forming what T. Gold called a
``fizzler'').  Ultimate formation of a neutron star would be
achieved only after angular momentum had been removed [perhaps
mainly into a surrounding disk and/or, as $P\rightarrow 10^{-3}$
seconds (the maximum spin rate of an axisymmetric neutron star),
through the transient formation of a Jacobi ellipsoid and subsequent
powerful gravitational radiation].  There is no obvious reason for
the final distribution of differential rotation after such a
genesis being the same as that of the 
proposed ``canonical'' CE from
the AIC of a magnetic WD with  
$B\approx 10^6$ G.  There are expected to be many
more WDs with smaller $B$ than those with $B\approx 10^6$ G.

(b) If the accreting WD's dipole $B$ greatly exceeds $10^6$ G,
the steady state $P$
of equation (\ref{PNS}) increases.
If any of these more slowly spinning
WDs were to collapse to a GRB source CE, that CE's spin
energy could not support
burst events with total energy near the maximum $10^{53}$ ergs.
However, if $\dot M$ could greatly exceed the Eddington limit
(indicated in Fig.~1) from a sufficiently massive Roche
Lobe overflow from the companion, $P(NS)\approx 10^{-3}$ seconds
might be achieved even from a $B\approx 10^9$ G WD, as
assumed by Usov in his CE model.

(c) The strongest sensitivity of CE emission pulse structures
would probably be to details of $B_p$, the initial polar field in
the newly formed differentially rotating neutron star, because of
possible magnification of small initial
$B_p$ differences to large ones
in the wound-up $B_\phi$. For example, $B_p$ details determine the
number density of toroids which  begin simultaneous wind-up
in different cylindrical regions around the spin-axis; these 
wound-up toroids overcome anti-buoyancy constraints and
break free at different times.  Locally, $B_p$ would be 
expected to change somewhat during these releases 
and any rewind-ups so CE emission
pulse shapes would not be repetitive during a GRB event.

\section{Discussion}

The required properties of GRB source CEs (summarized in
the Introduction) are total energy stored and emitted (a), 
peak power and fluctuations
within a given burst event (b), CE lifetimes (c), maximum
baryon loading in the CE emission (d), CE birthrate (e)
and very large variability among different CEs (f). 
None of these seem an embarrassing problem for the proposed model
CE genesis, structure and dynamics outlined in the Introduction
and described in Sections 2-9.  Indeed each seems a rather expected
consequence.  However, a crucial point which should
be considered further is the absence (so far) of any demonstrated
instability in the wind-up of the toroidal field for of order
$10^4$ turns (in about $10$ seconds) by the much more energetic
initial differential rotations in the neutron star.

A second related, but less crucial,
question is the robustness of our presumption
that during and after such toroidal wind up and release
the initial much smaller poloidal field component of the differentially
rotating neutron star is not hugely increased.  
If this does not turn out to be an adequate approximation, the
often observed sub-burst multiplicity could still come from
toroidal field wind-up and break-away in different cylindrical
regions with different wind-up times rather than from long time
delays for rewinding $B_\phi$.
Of course, 
because of the very great variability within the family of
GRB events, neither mechanisms may hold in all, or perhaps not
even in most cases, but at least one of them
should certainly not be uncommon.

Finally there is our unproven assumption % that there is enough non-axially
%symmetric magnetic field (or other interactions among $B_\phi$ in
%different places) in the toroidal field structures wound up by
that large toroidal field bundles wound-up by differential rotation
can overcome
anti-buoyancy restraints and break free as a unit (or almost so).
If, instead, buoyant toroidal field continually 
dribbled up and out to support
a steady state in which increasing $B_\phi$ from wind-up is balanced
by a that loss, there would be no strong fluctuations
in CE output.  Instead a CE would be an Usov-like pulsar with emission
decreasing monotonically after the first emission maximum is reached.
This is a generic problem for many kinds of CE models.
Why does the CE depart so far from a steady 
equilibrium that stored energy is released in huge sub-bursts (which
are often separated by
 very many characteristic engine periods) rather
than in smoother continuous steady way?  Here too such a question
needs further investigation.

\section{Acknowledgments}

We are happy to thank J. Applegate, E. Costa, C. Knigge,
J. Pringle, M. Rees, E. Spiegel, H. Spruit, and J. Stone 
for informative conversations, and 
the Institute of Astronomy (Cambridge) and the Aspen
Center for Physics for their hospitality while much of this
work was begun.

\clearpage

\begin{figure}
%\hskip 0.2 truein
\rotate{\psfig{file=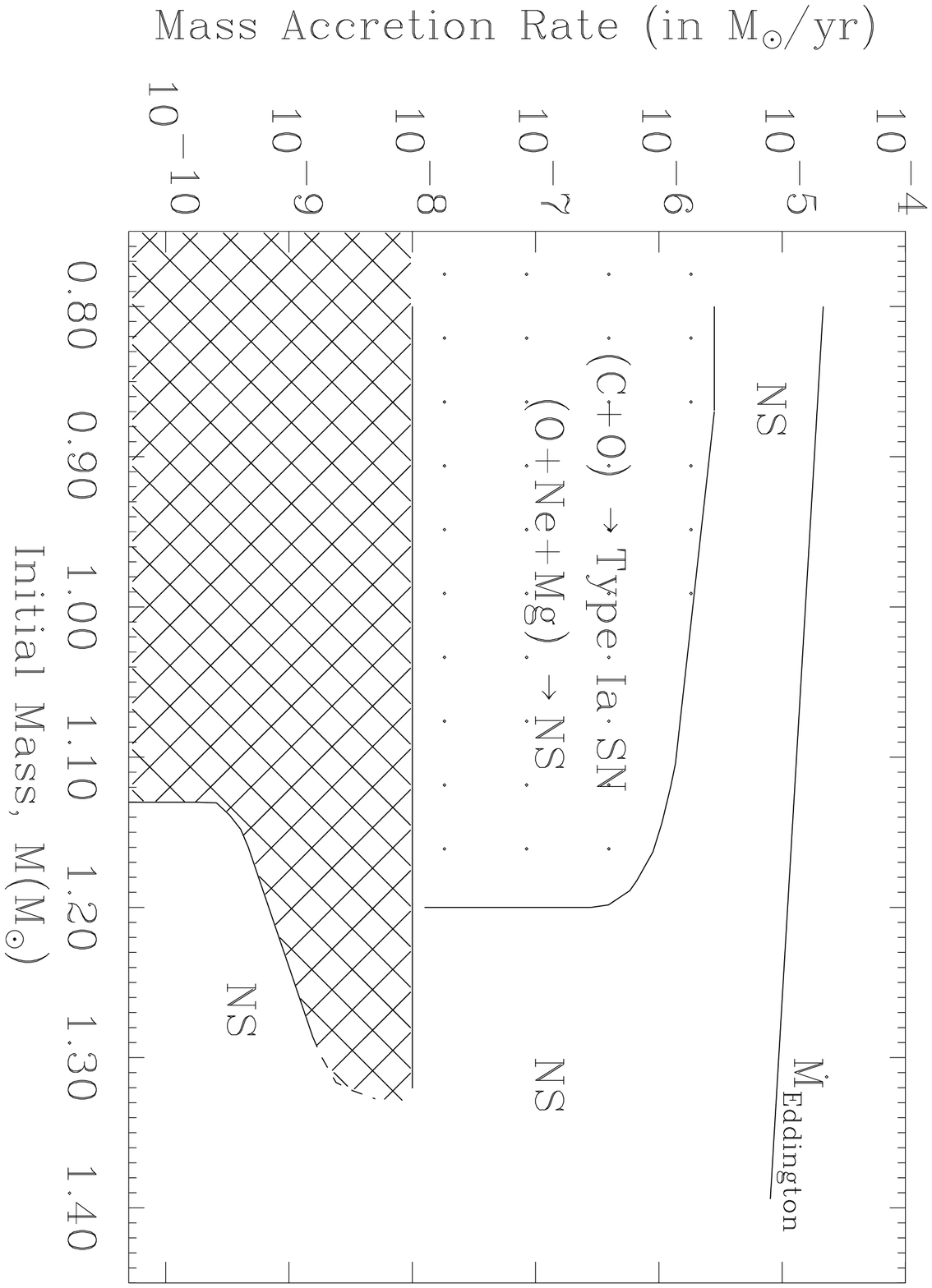,width=5.5truein}}
%\figcaption{Consequences of white dwarf accretion as
\caption{Consequences of white dwarf accretion as
a function of initial stellar mass ($M$) and assumed steady 
accretion rate ($\dot M$).  In the cross-hatched region, off-center
helium detonations keep the WD from ever gaining enough mass to
implode.  In the dotted region, the WD either becomes a Type Ia supernova
or a neutron star depending on its composition (see text). Figure based
upon Nomoto \& Kondo 1991.}
\end{figure}

%\begin{figure}
%\rotate{\psfig{file=density.ps,width=5.5truein}}
%\caption{Density profile of a $\gamma=4/3$ white dwarf (solid line)
%and of a
%cooled neutron star (dashed line).}
%\end{figure}

\begin{figure}
\hskip 0.5 truein
\psfig{file=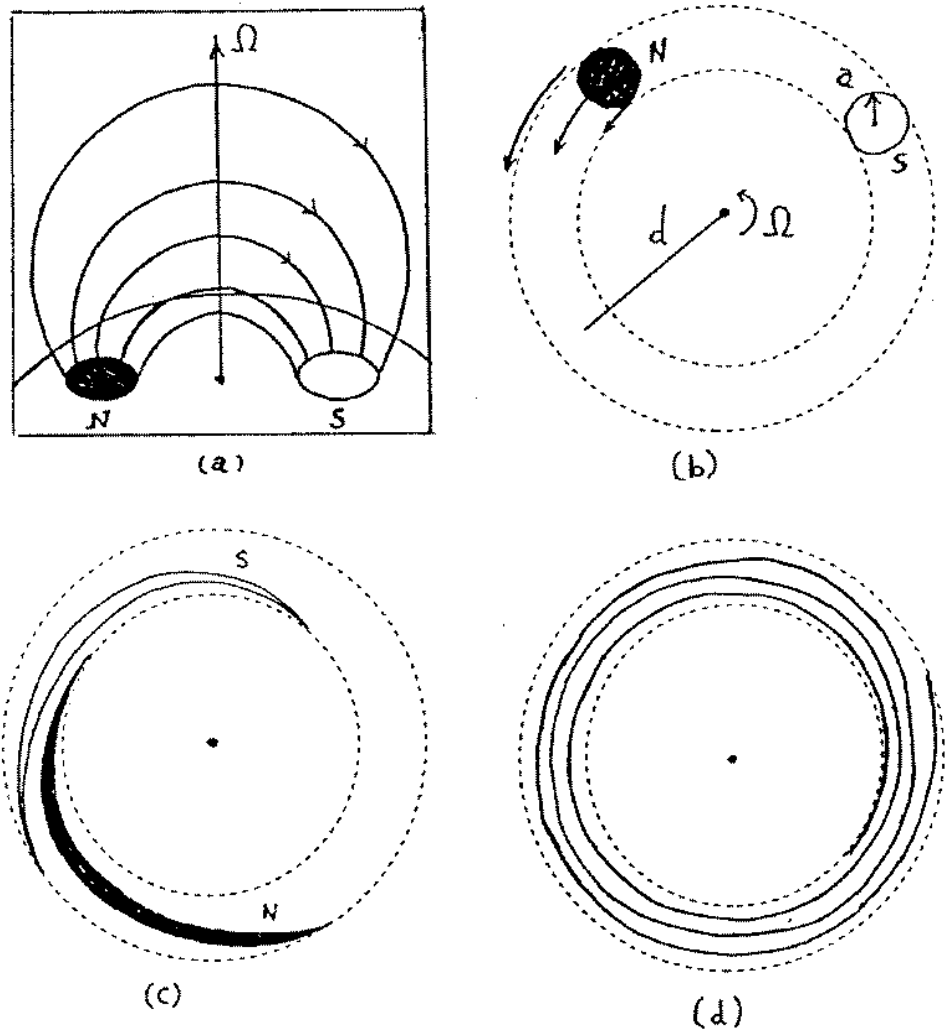,width=5.5truein}
\caption{Suppression of dipole and higher moments by surface shearing
from differential rotation.  Figure (a) shows a surface dipole
before shearing by differential rotation.  The filled circle is a
north-polar cap of radius $a$; the unfilled circle is a similar 
south-polar cap.
Figure (b) shows these same caps  viewed
along the spin-axis $\vec\Omega$ with the central dot the spot
where $\vec\Omega$ penetrates the surface.
Figure (c) shows the surface field after some differential rotation.
Figure (d) shows the north-polar cap 
after large differential rotation.  The south-polar cap (not shown)
would be another spiral falling in between the arms of the sheared
north-polar cap.  If the radii $a$ or distance $d$ were not 
identical for the two initial caps, the asymptotic state would be 
thin weakened north- and south-polar rings around $\vec\Omega$.}
\end{figure}

\end{document}